# Cable Capacitance Attack against the KLJN Secure Key Exchange

**Hsien-Pu Chen\*, Elias Gonzalez[†], Yessica Saez[†], and Laszlo B. Kish**

Department of Electrical and Computer Engineering, Texas A&M University, 3128 TAMU, College Station, TX 77843, USA; E-Mails: eliasg23@tamu.edu; yessica.saez@tamu.edu; Laszlo.Kish@ece.tamu.edu

[†]  These authors contributed equally to this work.

\*  Author to whom correspondence should be addressed; E-Mail: barrychen@tamu.edu

**Abstract:** The security of the Kirchhoff-law-Johnson-(like)-noise (KLJN) key exchange system is based on the Fluctuation-Dissipation-Theorem of classical statistical physics. Similarly to quantum key distribution, in practical situations, due to the non-idealities of the building elements, there is a small information leak, which can be mitigated by privacy amplification or other techniques so that the unconditional (information theoretic) security is preserved. In this paper, the industrial cable and circuit simulator LTSPICE is used to validate the information leak due to one of the non-idealities in KLJN, the parasitic (cable) capacitance. Simulation results show that privacy amplification and/or capacitor killer (capacitance compensation) arrangements can effectively eliminate the leak.

**Keywords:** KLJN; cable capacitance attack; capacitor killer; secure key exchange; unconditional security; privacy amplification.

## 1.  Introduction

The Kirchhoff-law-Johnson-(like)-noise (KLJN) key exchange system [1-4] was first introduced in 2005. Earlier, it was claimed that only Quantum Key Distribution (QKD) [5] could offer unconditional (that is information-theoretic) security. In due course, QKD's fundamental security claims have been debated by experts in the field [6-12]. Furthermore its practical realizations, including all commercial quantum communicators, have been fully cracked by hacking, that is, by utilizing non-ideal features of the hardware components [13-26]. While counter-measures were later proposed to overcome these attacks, when the idea of a new attack is unknown by the communicating parties and no counter-measures have been implemented yet, the eavesdropper can fully utilize such an attack [27-30].

Naturally, there have also been efforts to challenge KLJN's security [31-43]. Studies have consistently shown that both the ideal and the practical KLJN versions remain unconditionally secure [4,34-43] despite facing various attacks and related information leaks associated with the non-idealities of components in the system. The impacts of the attacks against practical KLJN systems have been weak.

With proper protocols, the Eavesdropper's (Eve's) probability of successful guessing of a bit can always be reduced to a value that is sufficiently close to 0.5 [3,34-38] to preserve unconditional security [4].

We will show that one of the most effective attacks against the practical KLJN system is the cable capacitance attack. It was first mentioned in 2006 [36], but it has never been tested. Subsequently, in 2008, a solution was suggested to eliminate this attack by adding a capacitor killer (capacitance compensation) arrangement [39].

In this paper, we use the industrial cable and circuit simulator LTSPICE by Linear Technology to simulate practical realizations of the KLJN system and to evaluate the cable capacitance attack. Solutions to mitigate this attack, such as the capacitor killer arrangement [39], and privacy amplification [44] are also tested.

## 2. The KLJN secure key exchange system

### 2.1 *The KLJN protocol*

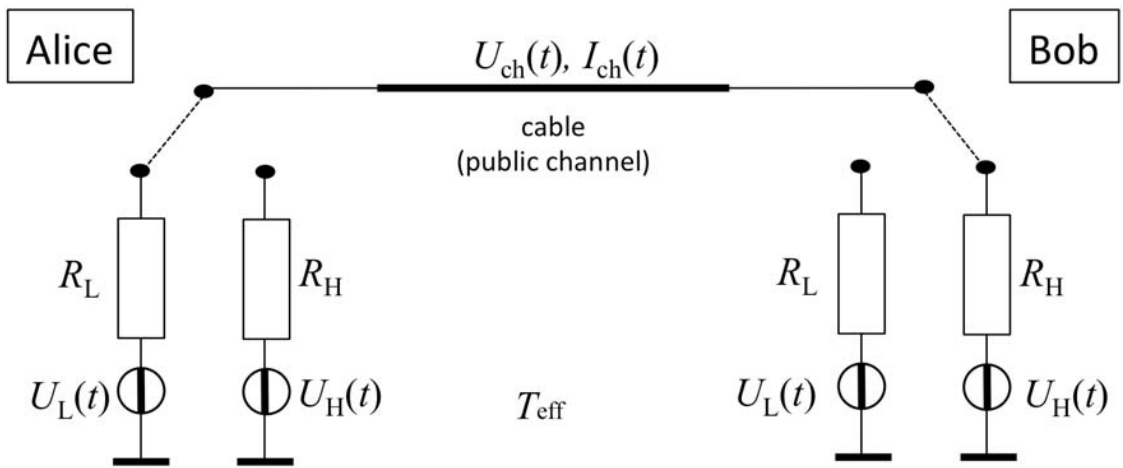

**Figure 1.** The core of the KLJN secure key exchange system [2]. The resistors values are $R_L$ and $R_H$. The thermal noise voltages, $U_L(t)$ and $U_H(t)$, are generated at an effective temperature $T_{eff}$ (typically $T_{eff} \geq 10^{14} \text{K}$) [40]. The channel noise voltage and current are $U_{ch}(t)$ and $I_{ch}(t)$, respectively.

The KLJN secure key exchange system [1-4,38-60] is based on Kirchhoff's Loop Law and the Fluctuation-Dissipation Theorem. The core KLJN system is illustrated in Fig. 1 [2]. It consists of a cable as an information channel, switches, and two identical pairs of resistors, $R_L$ and $R_H$, ( $R_L \neq R_H$ ), where $R_L$ represents the Low key bit (0) and $R_H$ represents the High key bit (1).

At the beginning of each bit exchange period (BEP), Alice and Bob randomly select $R_L$ or $R_H$ and connect the corresponding resistor to the cable. The Gaussian voltage noise generators in the figure represent either the Johnson noise sources of the resistors or external voltage noise generators emulating Johnson noise (filters are not shown). The noise is band-limited white noise with publicly agreed common bandwidth $B_{noise}$ and a publicly agreed common noise-temperature $T_{eff}$ [40]. The noises are statistically independent from each other and from the noise samples in the previous BEP [4]. Note, there are many advanced KLJN versions [41,42,56,60] with greater number of resistor values; some with different temperatures [56,60].

Within each BEP, Alice and Bob measure the mean-square channel noise voltage $\left\langle U_{ch}^2(t) \right\rangle$ and/or the channel noise currents $\left\langle I_{ch}^2(t) \right\rangle$ in the cable. The BEP has to be properly chosen to provide sufficient time for a good statistics of the mean-square noise voltages and currents but not enough time for Eve to effectively utilize possible information leaks due to hardware non-idealities. According to Johnson's noise formula:

$$\left\langle U_{ch}^2(t) \right\rangle = 4kT_{eff} \frac{R_A R_B}{R_A + R_B} B_{noise}, \tag{1}$$

$$\left\langle I_{ch}^2(t) \right\rangle = 4kT_{eff} \frac{1}{R_A + R_B} B_{noise}, \tag{2}$$

where $k$ is the Boltzmann's constant ($1.38 \times 10^{-23}\,\mathrm{J/K}$), $R_A$ and $R_B$ are the actual resistance values selected by Alice and Bob, respectively.

Based on equation 1 or 2, by measuring $\left\langle U_{ch}^2(t) \right\rangle$ and/or $\left\langle I_{ch}^2(t) \right\rangle$, and by knowing their own resistance value, Alice and Bob can determine [2] the resistor value at the other end and hence they can learn the bit value (0 or 1) there.

With the cable being public, an eavesdropper (Eve) can also measure the channel noise voltages and currents. If Alice and Bob use the same resistance values, so the arrangement is $R_L R_L$ or $R_H R_H$, the resulting noise levels are singular, (see Eqs. 1,2) thus the exchanged bit is non-secure and is discarded [2]. Conversely, the combinations $R_L R_H$ and $R_H R_L$ are degenerated because they produce the same noise levels. Thus the bit exchange is secure because Eve cannot differentiate between the two bit alternatives. From the noise levels (see Eqs. 1,2) Eve knows that Alice and Bob have exchanged a secure bit, but she does not know the location of $R_L$ and $R_H$.

In reality, the cable is non-ideal. Thus Eve can exploit the non-idealities of the cable, such as parasitic resistance, parasitic inductance and parasitic capacitance to attack the KLJN system.

## 2.2 Cable capacitance attack

In this paper, we assume coaxial cables because, in this case, the cable capacitance attack [36] can effectively be eliminated without the usage of privacy amplification. However, the attack works with any cable. Coaxial cables include two conductors: the inner wire, which is used as the KLJN channel, and the outer shield which is grounded (for the ground, see also Fig. 1). There is a non-zero capacitance between these two conductors that leads to capacitive currents. Part of the channel noise current is diverted by the parasitic capacitance, which causes a greater current at the end of the lower resistance. This gives Eve a chance to guess the value of the resistors with probability of success greater than 0.5.

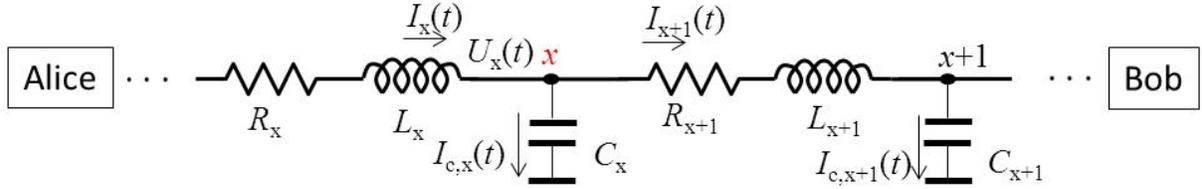

**Figure 2.** Cable model and cable capacitive currents.

Fig. 2, shows the distributed elements model of coaxial cables. According to Kirchhoff's current law, at position $x$, the channel noise current $I_x(t)$ is the sum of the capacitive current $I_{c,x}(t)$ through the parasitic capacitor element $C_x$, and the channel noise current $I_{x+1}(t)$. This is written as

$$I_x(t) = I_{c,x}(t) + I_{x+1}(t).$$  (3)

The capacitive current $I_{c,x}(t)$ is proportional to the time derivative of the channel noise voltage $U_x(t)$ and it is given by

$$I_{c,x}(t) = C_x \cdot \frac{dU_x(t)}{dt}.$$  (4)

We define the cross-correlation $\rho(x)$ [34] at position $x$ as the product of the channel noise current and the time derivative of the channel noise voltage:

$$\rho(x) = \left\langle I_x(t) \cdot \frac{dU_x(t)}{dt} \right\rangle_\tau,$$  (5)

where $\langle \ \rangle_\tau$ means finite time ($\tau$) average. The location-dependence of $\rho(x)$ represents information leak [34].

## 3. Realization of the attack

The cable and a circuit simulator LTSPICE by Linear Technology was used to emulate the practical KLJN system with the RG58 coaxial cable from its library. Throughout the simulations, we assumed that Alice selected $R_L = 1$ kohm and Bob $R_H = 9$ kohm, see Fig. 3.

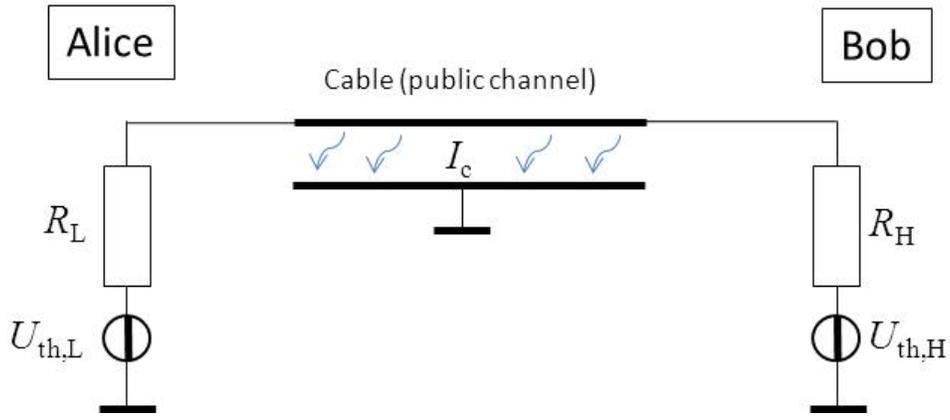

**Figure 3.** The simulated KLJN system with capacitive current $I_c$. The generator voltages $U_{th,L}$ and $U_{th,H}$ are the Johnson noise voltages of $R_L$ and $R_H$, respectively.

### 3.1 Generating the noise

For the simulations, we generated Gaussian band-limited white noises. According to Johnson's noise formula, the required rms noise voltage $U_{th}$ is

$$U_{th} = \sqrt{4kT_{eff}RB_{noise}} \ .$$

(6)

As the mean value is zero, the rms noise voltages are the same as their standard deviations (denoted as $\sigma_L$ and $\sigma_H$ for $U_{th,L}$ and $U_{th,H}$, respectively). Thus

$$U_{th,L}/U_{th,H} = \sigma_L/\sigma_H = \sqrt{R_L/R_H} \ ,$$

(7)

where $\sqrt{R_L/R_H} = \sqrt{1/9}$, thus $\sigma_L/\sigma_H = 1/3$. For the simulations, the rms thermal noise voltages of $R_L$ and $R_H$ were chosen as 1V and 3V, respectively, corresponding to $T_{eff} \approx 7 \times 10^{16}$ K.

Fig. 4(a) shows the probability density function (histogram) of the noise voltage of $R_L$. In Fig. 4(b) the cumulative distribution as normal probability plot can be seen where a straight line indicates exact normal distribution.

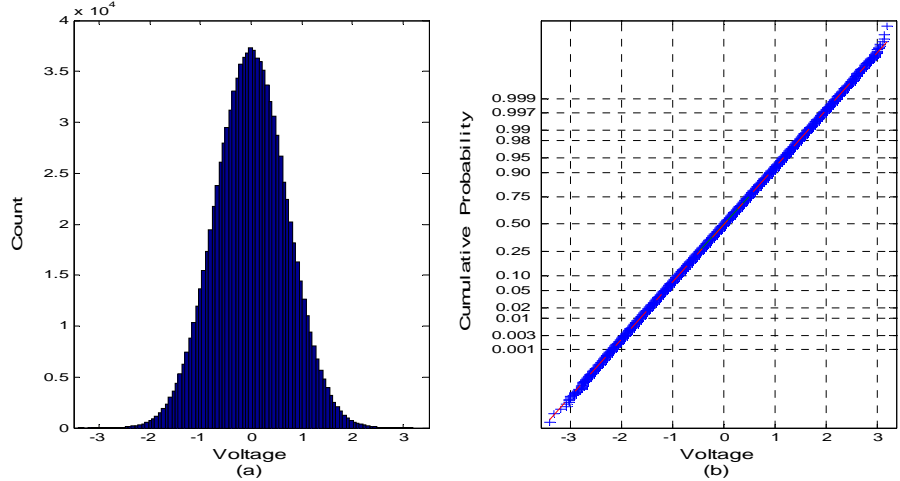

**Figure 4.** Statistics of the Johnson noise voltage of $R_L$ with $10^6$ samples. (**a**) probability density function (histogram); (**b**) cumulative distribution as normal probability plot.

### 3.2 Comparing a lumped and the distributed element models at different wavelengths

First, for enhanced computational speed, we explored the possibility of using lumped element cable model for the simulations because the continuum model simulations are at least 1000 times slower. Our data below proves that lumped elements can be used for high-accuracy simulations at the operational conditions of KLJN.

The quasi-static condition is required for the security of the KLJN system [2,34]. That means

$$L_{ch} \ll \lambda = c/B_{noise} \qquad \text{or} \qquad \gamma = \lambda/L_{ch} \gg 1, \qquad (8)$$

where $L_{ch}$ is the cable length, $\lambda$ is the shortest wavelength at the highest frequency component of the noise bandwidth $B_{noise}$, $c$ is the propagation velocity in the cable, and $\gamma$ is the ratio of the wavelength to the cable length. It has been assumed that $\gamma$ must be at least around 10 to fulfill the KLJN conditions [34,49,50,57,58] (i.e., approximate quasi-static electrodynamics; see [49,50] about the proof that there are no waves in this limit).

Fig. 5(a) and 5(b) shows the simple lumped element model and the distributed model of the RG58 coaxial cable, respectively. Based on the specific inductance and capacitance, the propagation velocity $c$ in the RG58 coaxial cable is $2 \times 10^8$ meter/sec. Three simulations were run to compare the resultant voltage waveforms at Alice's side, at three different noise bandwidths $B_{noise}$ (250 kHz, 25 kHz, 0.25 kHz) on these 2 models. The cable length was set at 1000 meters, based on equation 8, the three corresponding wavelengths ($\lambda$) were 800 m, 8 km, and 800 km, while the corresponding $\gamma$ ratios were 0.8, 8 and 800. Other parameters such as the component values of the models used in the simulations are also shown at Fig. 5.

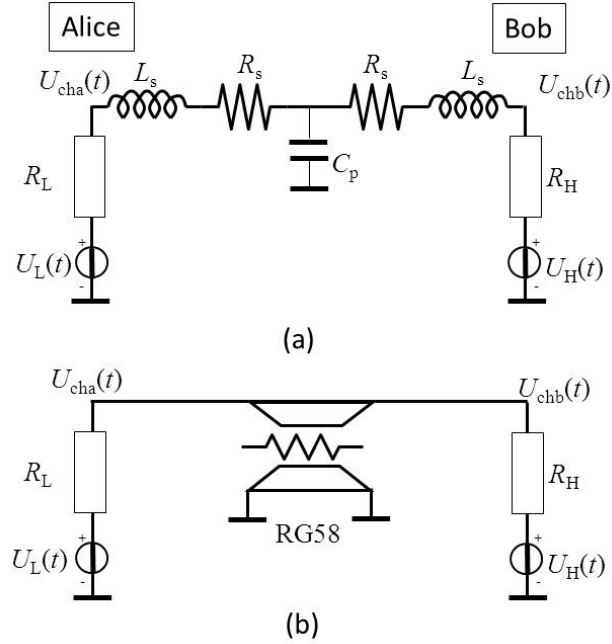

(a)

(b)

**Figure 5.** The RG58 coaxial cable models (1000 m length) with $R_L$ (1 kohm) and $R_H$ (9 kohm).

**(a)** The lumped element model: the component values: $R_S = 10.5 \, \text{ohm}$, $L_S = 125 \, \mu\text{H}$, $C_P = 100 \, \text{nF}$.

**(b)** The distributed model had the following parameters: $R = 0.021 \, \text{ohm/meter}$, $L = 250 \, \text{nH/meter}$, $C = 100 \, \text{pF/meter}$. The characteristic impedance of the cable is 50 ohms.

Fig. 6 shows the simulation results, where $U_{cha,lump}$ and $U_{cha,dist}$ are the voltage timefunctions of the lumped and distributed element models, respectively. In Fig. 6(a), the two waveforms are significantly different for the shortest wavelength with $\gamma = 0.8$. In such a case, the waves can only be simulated with the distributed model. However, this situation is irrelevant for the operation of KLJN, as mentioned above.

In Fig. 6(b), with $\gamma = 8$, the two waveforms are very similar whereas in Fig 6(c), at $\gamma = 800$, the two waveforms are indistinguishable. Thus we can conclude that for situations $\gamma \geq 8$, the lumped element simulations are satisfactory. Both cases are fine for the KLJN operation and we will use the $\gamma \geq 800$ condition in the rest of the paper.

For our resistor values $R_{\mathrm{L}} = 1$ kohm and $R_{\mathrm{H}} = 9$ kohm, the cut-off frequency by the cable capacitance is 1.76 kHz and 17.6 kHz for a 1000 and a 100 meters cable, respectively. To avoid that the cable capacitance truncates the effective bandwidth of the noise, we used noise bandwidth $B_{\mathrm{noise}} = 0.25$ kHz for the noise generators ($\gamma = 800$ at 1000 meters and $\gamma = 8000$ at 100 meters).

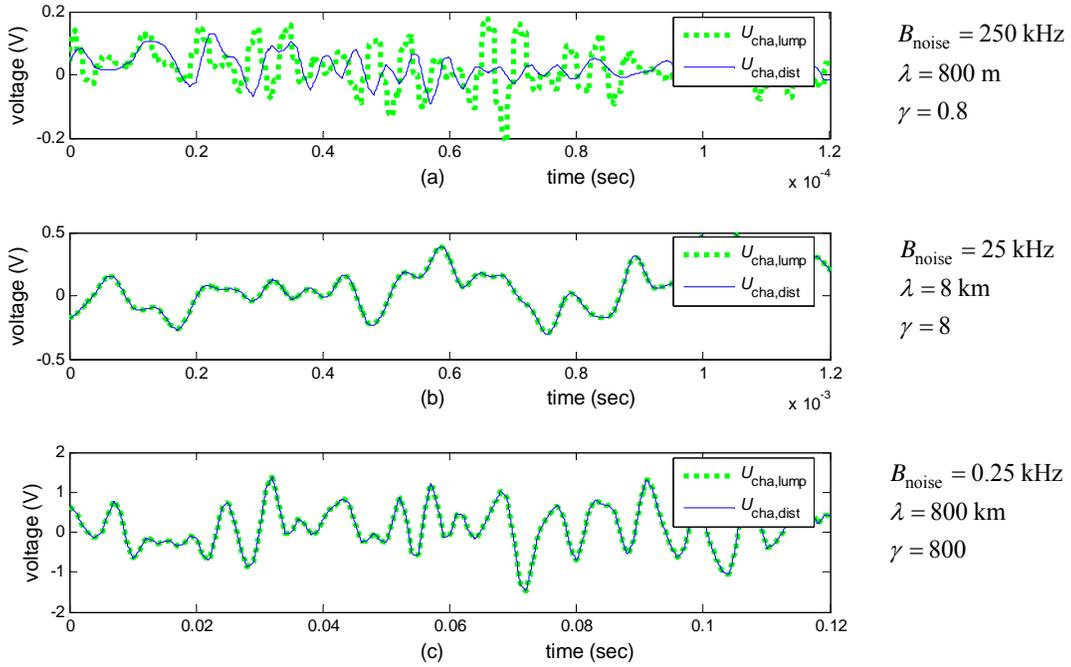

**Figure 6.** The voltage waveforms at Alice's side, $U_{\mathrm{cha,lump}}$ and $U_{\mathrm{cha,dist}}$, for the lumped and distributed element models, respectively, for a 1000 meters cable, at **(a)** $\gamma = 0.8$; **(b)** $\gamma = 8$; **(c)** $\gamma = 800$.

*3.3 The attack protocol*

In this section, we discuss the information leak caused by the cable capacitance and evaluate Eve's success probability of guessing the key bits. The fixed bit arrangement is used between Alice and Bob.

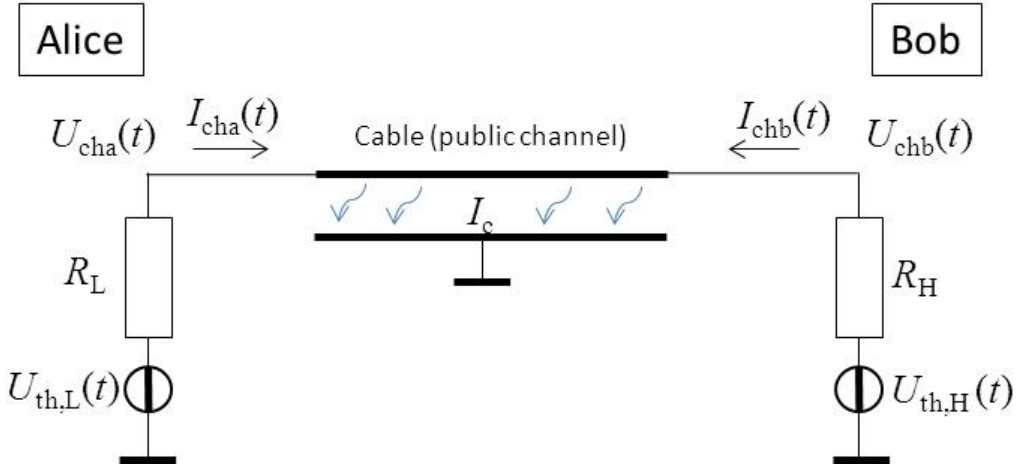

**Figure 7.** The simulated model with LH bit arrangement ( $R_L$ = 1 kohm and $R_H$ = 9 kohm). $U_{cha}(t)$, $I_{cha}(t)$, $U_{chb}(t)$ and $I_{chb}(t)$ are the voltages and currents at Alice's and Bob's ends, respectively.

During the exchange of the *i*-th bit, Eve measures the cross-correlations

$$\rho_{ia} = \left\langle I_{cha}(t) \cdot \frac{dU_{cha}(t)}{dt} \right\rangle_{\tau},$$ (9)

$$\rho_{ib} = \left\langle I_{chb}(t) \cdot \frac{dU_{chb}(t)}{dt} \right\rangle_{\tau},$$ (10)

where $U_{cha}(t)$, $I_{cha}(t)$, $U_{chb}(t)$ and $I_{chb}(t)$ are the channel voltages and currents at Alice's and Bob's ends, respectively, see Fig. 7. The time average $\langle \ \rangle_{\tau}$ is taken over the bit exchange period $\tau$. Eve calculates $\rho_i = \rho_{ia} - \rho_{ib}$ $(i = 1,...,N)$ and decides as follows:

$$
\begin{array}{ll}
\text{If} \quad \rho_i > 0 \quad \text{then} \quad q_i = 1 & \text{(\textit{Eve guessed the bit correctly})} \\
\text{If} \quad \rho_i < 0 \quad \text{then} \quad q_i = 0 & \text{(\textit{Eve guessed the bit wrongly})}
\end{array}.
$$ (11)

When *N* approaches infinity, the probability of Eve's successful guessing of the bits is equal to the expected value of *q* and

$$\langle q_i \rangle_N = p_E = 0.5 + \varepsilon, \text{ where } 0 \le \varepsilon < 0.5,$$ (12)

where non-zero $\varepsilon$ represents an information leak. When $\varepsilon = 0$ the KLJN key exchange system is perfectly secure. We found that the higher the difference between the resistances, higher the bandwidth, or higher the parasitic capacitance (longer the cable), the higher the leak.

### 3.4 Simulation results of the cable capacitance attack

We simulated 6 different attack scenarios with these parameters: $R_L = 1$ kohm, $R_H = 9$ kohm, noise bandwidth $B_{noise} = 0.25$ kHz, sampling period $t_s = 1$ msec; for 3 different single-bit exchange durations (measured by the unit of the autocorrelation time of the noise), 20, 50, 100; at 2 different cable lengths, 100 and 1000 meters. At each scenario, the key was 1000 bits long.

The simulation results are shown in Table 1. At bit exchange duration = 20 (50 bits per second), with a 100 meters cable, Eve's success rate was 50.9%. However, when the cable length was increased to 1000 meters with the other parameters unchanged, Eve's success rate increased to 62.2%.

**Table 1.** Attack simulation results - Eve's success rate $p_E$ (%) with 1000 bits key length

| Bit exchange duration | Bits per second | 100 meters cable | 1000 meters cable |
|:---:|:---:|:---:|:---:|
| 20 | 50 | 50.9% | 62.2% |
| 50 | 20 | 52.1% | 69.7% |
| 100 | 10 | 52.6% | 76.9% |

When the bit exchange duration was increased to 50 and 100, Eve's success rate increased accordingly as shown in Table 1. In the most effective attack case, Eve success rate was 76.9%.

## 4. Defense against the attack

### 4.1 Capacitor killer

The parasitic capacitance of the RG58 coaxial cable can be eliminated by the well-known capacitance compensation technique, called capacitor killer arrangement [39], providing the same voltage on the outer shield of the cable as on the inner wire. This can be done by an ideal voltage follower, see Fig. 8. There is no capacitive current from the inner wire to the outer shield thus the attack is nullified.

We simulated the capacitor killer arrangement at the most effective attack scenario (i.e., when Eve success rate was 76.9%). The simulation results showed that Eve success rate was reduced from 76.9% to 50.1%. This indicated that the capacitor killer is very effective in eliminating the leak due to the parasitic capacitance at the practical cable conditions we tested.

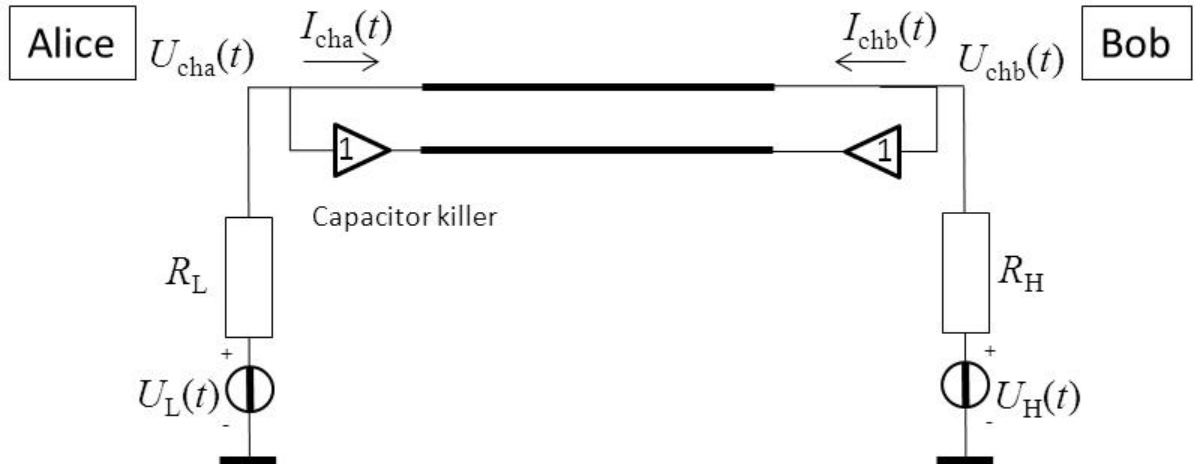

**Figure 8.** The KLJN system with the capacitor killer. An ideal voltage follower is driving the outer shield, which is not grounded at this time.

### 4.2 Privacy Amplification

Another method to secure the key exchange and to reduce information leak is by utilizing privacy amplification [44]. Due to the extraordinarily low bit error probability of the KLJN system [51-53], privacy amplification (which is basically an error enhancer) can be used to effectively reduce any information leak. The simplest and most secure concept [44] is that Alice and Bob XOR the subsequent pairs of the key bits (i.e., XOR the first and the second bits to get the first bit of the new key, XOR the third and the fourth bits to get the next one, etc.). In this way the length of the new key will be half of the original one but Eve's success probability will get closer to 0.5; that is, it moves toward the limit of zero information. We simulated the effect of this technique by utilizing the most effective attack scenario (see Table 1). The simulation results showed that by XOR-ing once, Eve's success probability was reduced from 76.9% to 64.2%, which was further reduced to 54.4% by XOR-ing a second time resulting a cleaner key with the corresponding significantly higher security and one quarter of its original length.

### Conclusions

By utilizing the LTSPICE simulator we have validated the cable capacitance attack. Both the capacitor killer method and privacy amplification have been able to eliminate the attack. The unconditional security of a practical KLJN key exchange system [4] has been preserved against this attack, too.

Note that the temperature compensation method [59] based on the non-equilibrium thermodynamical aspects of KLJN to eliminate the information leak at wire resistance attack, does not reduce the efficiency of the cable capacitance attack.

Finally, we mention that there is a new, advanced protocol, the Random-Resistor-Random-Temperature (RRRT) KLJN scheme [60], where all the former attacks become invalid or incomplete, and currently no known attack works against it. This is true also for the cable capacitance attack presented above: it is invalid against the RRRT-KLJN scheme. Further studies will be needed to find ways for all the former attack schemes to successfully extract information from the RRRT-KLJN system [60] at non-ideal conditions where they may leak information.